# 1. Chemistry of Low Mass Substellar Objects

**Katharina Lodders & Bruce Fegley, Jr.**

**Abstract**: "Brown dwarfs" is the collective name for objects more massive than giant planets such as Jupiter but less massive than M dwarf stars. This review gives a brief description of the classification and chemistry of low mass dwarfs. The current spectral classification of stars includes L and T dwarfs that encompass the coolest known stars and substellar objects. The relatively low atmospheric temperatures and high total pressures in substellar dwarfs lead to molecular gas and condensate chemistry. The chemistry of elements such as C, N, O, Ti, V, Fe, Cr, and the alkali elements play a dominant role in shaping the optical and infrared spectra of the "failed" stars. Chemical diagnostics for the sub-classifications are described.

## 1.1. Introduction

This year marks the 10$^{th}$ anniversary of the first discovery of Gl 229B, a bona-fide brown dwarf – a substellar object too low in mass to sustain hydrogen fusion to shine as a star (Nakajima et al. 1995, Oppenheimer et al. 1995). In the 1960s, it was already recognized that a lower mass limit of $\geq$0.08 solar masses is required for dwarf stars to enter the main sequence and that such objects only release energy from gravitational contraction and evolve into "black dwarfs" (Kumar 1963). These low-mass objects became more colorful in 1975, when Tarter called them "brown dwarfs", a name still commonly used. However, since these low-mass and faint objects appear reddish purple to magenta instead of brown (Burrows et al. 2001), we use "low mass objects", "substellar objects" or "substellar dwarfs" in this review.

Since 1995, a plethora of objects less massive and cooler than true M dwarf stars, but more massive than giant planets such as Jupiter, has become known through the Two Micron All Sky Survey (2MASS), Deep Survey of the Southern Sky (DENIS), and Sloan Digital Sky Survey (SDSS) infrared sky searches. These objects extend the spectral sequence of stars to fainter and lower mass objects (e.g., reviews by Chabrier & Baraffe 2000, Basri 2000, Burrows et al. 1999, 2001). By 1999 two new classes, L and T, were added to the stellar spectral classification (Kirkpatrick et al. 1999, Martin et al. 1999. Current counts are ~300 L dwarfs, and ~60 T dwarfs[1].

---

[1] see http://spider.ipac.caltech.edu/staff/davy/ARCHIVE ;
http://www.astro.ucla.edu/~adam/homepage/research/tdwarf/ and
http://www.astro.ucla.edu/~mclean/BDSSarchive/

The L dwarfs more closely resemble M dwarfs in their spectra and include the lightest real stars and the heaviest substellar objects. The T dwarfs, characterized by methane and water absorption bands, have closer spectral resemblance to giant planets but are much more massive than ~13 Jupiter masses, which is the mass limit for deuterium burning (see below). The latest suggestion for addition to the spectral sequence is the letter "Y" for objects that lack water absorption bands because the outer atmospheres of Y dwarfs should be cool enough (<~500 K) to sequester water into clouds (Ackerman & Marley 2001, Burrows et al. 2003). There are no direct observations yet of objects that are cooler than the coolest known T dwarfs ($T_{eff}$ ~700 K) to complete the bridge to Jupiter ($T_{eff}$ ~125 K).

The detection of the dim substellar objects is challenging because they are small (about Jupiter's radius) and low in mass. The upper mass limit for substellar dwarfs is about 7-8% that of the Sun (for comparison Jupiter' mass is 0.095%), insufficient to sustain H-burning like in real stars. Substellar objects may burn deuterium if they exceed ~13 Jupiter masses and burn $^6$Li and $^7$Li if they exceed 65 Jupiter masses (the mass limits for H and D burning are for objects with solar metallicity, see Burrows et al. 2001). In the most massive substellar objects, D-burning lasts up to ~$3\times10^7$ years and Li-burning up to ~$3\times10^8$ years, whereas H-burning operates several $10^9$ years in dwarf stars. Most of the energy released by a substellar object over its lifetime is from gravitational energy gained during its formation and contraction as it sits in space and cools.

Substellar dwarfs never exceed ~3000 K near their surfaces. As blackbody curves show, the more an object cools, the less it is visible at optical wavelengths. M dwarf stars emit most strongly at red optical wavelengths (~0.75 μm) but maximum emissions of the cooler L (~1200 < $T_{eff}$ < 2200 K) and T dwarfs (300 K < $T_{eff}$ < ~1200 K) are shifted to longer wavelengths in the near infrared (1–2 μm). Substellar objects also appear intrinsically fainter (i.e., compared to an M dwarf star at the same distance) because of the significant decrease in the energy density at maximum in cooler blackbodies. Furthermore increased opacity from metal oxides, metal hydrides, water, carbon monoxide, methane, and ammonia at different near-infrared wavelengths may reduce the overall energy flux by so much that essentially nothing of an ideal blackbody intensity curve remains.

The molecular chemistry in the relatively dense and cool atmospheres of substellar objects leads to significant changes in the observed far-red and near-infrared spectra when compared to hotter dwarf stars. Chemical thermodynamics and kinetics determine the gases present at a given depth in the atmosphere as function of temperature, pressure, and overall elemental composition. The chemical speciation changes when a solar composition gas is cooled from high to low temperatures and is largely responsible for the observed differences in stellar and sub-stellar spectra. At high photospheric temperatures such as in K and early M dwarf stars, monatomic ions and neutral atoms dominate, but at the lower temperatures near the M/L transition and in T dwarfs, neutral atoms, particularly of the alkali elements, and molecules such as TiO, VO, FeH, CrH, CO, $H_2O$, and $CH_4$ become more and more important. Condensation of some elements from the gas into clouds also occurs in L and T dwarfs.

## 1.2. Classification schemes

The spectral M, L, and T dwarf classifications make use of the temperature dependent appearance and disappearance of ions, atoms, and molecules in the optical and near IR spectra of the these objects e.g., Kirkpatrick et al. 1999, 2000 (=K99, K00), Martin et al. 1999 (=M99), Burgasser et al. 2002a (=B02), 2003, Geballe et al. 2002 (=G02), Leggett et al. 2002 (=L02), McLean et al. 2003 (=Mc03), Nakajima et al. 2004). Some of the major characteristics also used for classification are summarized in Table 1.

Spectral subtype numbers from zero for the hottest (e.g., L0, T0) to (currently) nine for the coolest (e.g., L9, T9) are set when certain spectral features appear or disappear (Table 1). Note that currently there are 2 scales for subtyping the L dwarfs that differ in the assignments of the higher L subtype numbers (M99, K99, K00). Here we follow the majority of the literature and use the Kirkpatrick et al. scale; however, the conversion of this subtype scale to that by Martin et al. is also given in Table 1.

Late M dwarfs (>M6) are characterized by increasingly stronger TiO, VO $H_2O$, FeH, CrH, CaH, and MgH bands. Water bands grow stronger from mid-M dwarfs through the L and T dwarf sequence in all spectral ranges. The L dwarf sequence starts when TiO and VO bands begin to weaken while CrH, FeH, and water bands strengthen in the optical. In the near infrared, CO, CrH and FeH absorptions strengthen towards the mid-L subclasses, and then decline.

However, the FeH absorption at 0.9896 μm reappears and strengthens again in early T dwarfs and reaches a maximum in strength around T5 before completely disappearing at T8 (Burgasser et al. 2002a,b). The band of CrH at 0.9979 μm behaves similarly and could be a misidentified FeH band according to Cushing et al. (2003) who investigated and identified many FeH transitions. However, the assignment of the band at 0.9979 μm to CrH is confirmed by opacity computations for CrH by Burrows et al. (2002a) and FeH by Dulick et al. (2003). These studies also show that the FeH/CrH abundance ratio plays a critical role for the observability of CrH.

Table 1. Spectral Classification of L and T Dwarfs

| Spectral type M99 | K99 | Spectral characteristics | Chemistry |
|---|---|---|---|
| M8 | **M8** | TiO (705.3, 843.2 nm) at max. (M99, (K99)) | TiO at max. |
| M9 | **M9** | VO (740, 790 nm) at max (M99, (K99)) | VO at max |
| L0 | **L0** | blueward TiO gone, redder TiO remains (K99,M99); Al I, Ca I disappear in J band (Mc03) ; Rb I (780.0, 794.8 nm); Cs I (852.1, 894.3 nm) weak (K99, M99) | Ca, Al, Ti cond. |
| L1 | **L1** | VO bands (740, 790nm) weaken; Na I (589.0 & 589.6 nm) doublet weakens; Rb I, Cs I increasing in strength (K99,M99) | V cond $Cs > Cs^+$ |
| L2 | **L2** | all TiO bands gone except at 843.2 nm (K99); Ca I triplet still in K band (Mc03) | |
| L2.5 | **L3** | Fe I at 1.189 μm still present (Mc03) | |
| | **L3.5** | Fe I at 1.189 μm gone (Mc03); K I (1.169, 1.197, 1.25μm) peak in strength (R01) | Fe cond. |
| L3 | **L4** | all VO gone (K99); CrH (861.1nm) = FeH (869.2nm); FeH at 869.2 nm & J band at max. (R01, B02, K99, M99, Mc03); K I doublets in J-band at max (B02, Mc03) | FeH ≈ CrH K at max. |
| L4 | **L5** | $CH_4$ appears at 3.3 μm (Noll et al. 2000); CaH at 685 nm still seen (M99, Mc03); FeH (869.2 nm & J band) weakens (B02,K99,M99); CrH 861.1 nm at max. (B02, K99, M99); optical Rb I, Cs I strengthening, but weaker in J band (K99, Mc03) | |
| L5 | **L6** | CrH weakens in optical (K99, M99) Li I (670.8nm) still observed in some objects (K99, M99) | Cr cond. Li = LiCl |
| L5.5 | **L7** | TiO 843.2nm (K99, M99) essentially absent, ditto TiO 549.7, 559.7, 615.9, 615.9, 638.4nm (M99); Li I most likely absent (K99,M99) | |
| L6 | **L8** | subtle $CH_4$ band in K band (G02, N04); no or subtle $CH_4$ in H band (B02,G02,Mc03); Na I (589.0, 589.6 nm) very broad, Na I barely seen in optical and J band (K99,M99,Mc03); K I doublets at min. in J band (N04) | CO=$CH_4$ |
| | **T0** | $CH_4$ appears in K and H bands (G02) FeH (989.6 nm) weaker (N04) or as strong as in L8 (B02b) | CO=$CH_4$ |
| | **T1** | weak $CH_4$ in H, J, & K bands (B02,G02); strengths of CO=$CH_4$ in K band; K I (1.2432, 1.2522μm) present in J band (B02) | CO=$CH_4$ |
| | **T2** | strengths of $CH_4$>CO in K band (B02); CrH (861.1 nm) & FeH (869.2nm) disappear (B03); FeH (989.6 nm) weaker (N04) or as strong as in L8 (B02b); Cs I (852.1, 894.3nm) peak in strength (B03) | |
| | **T3** | CO (2.295μm) very weak or absent in K band (B02, Mc03,N04) | |
| | **T4** | FeH (989.6nm), CrH (996.9nm) increase again in strength (B02,B02b,B03,N04) | |
| | **T5** | No CO in K band (B02); FeH (989.6nm) & CrH (996.9nm) peak in strength (B02b, B03,N04); K I at maximum in J band (B02, Mc03,N04) | $Na_2S$ cond. |
| | **T6** | K I at in J band begins to weaken (B02) | . |
| | **T7** | K I in J band very weak (B02, Mc03) | |
| | **T8** | no FeH (989.6nm) (B02, Mc03); no K I in J band (B02, Mc03); Cs I marginally detectable in optical (B03) | ($NH_3$=$N_2$)? |

B02: Burgasser et al. 2002a,b; G02: Geballe et al. 2002; K99: Kirkpatrick et al. 1999, 2000; M99: Martin et al. 1999; Mc03: McLean et al. 2003; N04: Nakajima et al. 2004, R01: Reid et al. 2001

H band centered at ~1.6 μm; J band centered at ~1.25 μm; K band centered at ~2.2 μm

The L dwarfs show CO overtone bands in the 1–2.5 μm range and the band at 2.3μm is traceable into the T dwarf sequence up to T3 (B02, Mc03). The strongest band of $CH_4$ at 3.3 μm already appears in L5 dwarfs (Noll et al. 2000), and in some L6.5 and L8 dwarfs very weak bands of $CH_4$ are seen at 2.2 and 1.6 μm (see Nakajima et al. 2004, McLean et al. 2003). The T type is defined by the onset of methane absorption in the 1.6 and 2.2 μm photometric bandpasses (H and K bands). This causes the shift towards blue in the infrared J-H and H-K colors in the spectral sequence near the L/T transition (L02, Golimowski et al. 2004). Within the T sequence increasingly stronger $H_2O$ and $CH_4$ absorptions appear between 1.0–2.2 μm (hence also the name "methane" dwarfs), and in addition, optical spectra of T dwarfs are shaped from collisionally induced absorption (CIA) by $H_2$ (K99, Tokunaga & Kobayashi 1999, Liebert et al. 2000, B02, G02).

Throughout the L-sequence, the absorption resonance doublets of monatomic Na I (0.5890, 0.5896μm) and K I (0.7665, 0.7699μm) weaken and become strongly pressure broadened (see, K99, Tsuji et al. 1999, Burrows et al. 2000, 2002b, Liebert et al. 2000, Burrows & Volobuyev 2003). This continues within the T dwarf sequence, where the extremely broadened wings of the Na I and K I doublet absorptions dominate the slope of large portions in the red and near IR spectra so that most of the sub-typing of T dwarfs is done better with near infrared spectra (e.g., B02, G02). The near infrared K I doublets in the J band (Table 1) peak in strength around L3.5-L4 (Reid et al. 2001, B02, Mc03) and then decline but reach a second increase in strength around T5/T6 (e.g., B02), a behavior that is mimicked by FeH noted above.

The lines of Rb I (0.7800, 7948μm) and Cs I (0.8521, 8943μm) appear near the M/L transition, grow in strength throughout the L sequence, and reach a peak in strength at T2, after which their strength decreases (K99, B02). The presence of the resonance doublet of monatomic Li I (0.6708μm) - a useful but limited test for the substellar nature of an object (see below) - appears to be restricted to earlier L subtypes of up to L6.

The introduction of the L and T classes is relatively new, and the definitions of the L and T subtypes are still work in progress. Characterization of the subtypes depends on the number of known and observed substellar dwarfs, the spectral wavelength range (optical vs. infrared) covered and resolution of the spectra, and the choice of spectral features used to parameterize the subclasses. Kirkpatrick et al. (1999, 2000) and Martin et al. (1999) utilized red optical spectra (0.63-1.01μm) for M and L dwarfs, and near-infrared spectral features were used to classify T dwarfs (BO2, G02). Work on linking the optical and infrared observations into one scheme is ongoing and unified classification schemes for L and T dwarfs using spectral indices from the same wavelength ranges are under development (see, e.g., M99, K99, K00, Reid et al 2001, Tokunaga & Kobayashi 1999, Testi et al. 2001, B02, G02, Mc03, Nakajima et al. 2004). Geballe et al. (2002) combined optical (0.6-1.0 μm) and infrared (1.0-2.5 μm) spectra for M, L, and T dwarfs to find suitable infrared flux indices that also can be linked to the optical classification scheme of L dwarfs from K99 and M99. Apparently only the water band at 1.5 μm is useful for sub-typing the entire L and T sequence and for relating the infrared and optical classification schemes (but see also Tokunaga & Kobayashi 1999, Reid et al. 2001, Testi et al. 2001, Knapp et al. 2004). For objects of higher spectral type than L3 in the schemes of M99 and K99, the strength of the methane band at 2.2 μm provides an additional common index for the L and T classes (G02). McLean et al. (2003) suggested a classification scheme for L and T dwarfs based on several indicators such as the relative strengths of the atomic lines of Na, K, Fe, Ca, Al, and Mg and bands of water, carbon monoxide, methane, and FeH in near infrared spectra.

## 1.3. Effective temperatures along the L and T dwarf sequence

The effective temperatures ($T_{eff}$) of substellar objects depend on their total mass, radius, and their age, and are generally below 3000 K (see Burrows et al. 2001). The effective temperatures are related to luminosity (L) by $L = 4 \pi R^2 \sigma T_{eff}^4$, where R is the radius of a substellar dwarf and σ the Stephan-Boltzmann constant. The luminosities can be obtained from the bolometric magnitudes if bolometric (filter) corrections as well as parallaxes of the objects are known. Bolometric corrections were recently computed for a large set of L and T dwarfs (Golimowski et al. 2004) which combined with parallax measurements (Dahn et al. 2002, Vrba et al. 2004) and estimates of substellar object radii give $T_{eff}$ estimates summarized in Table 2.

Other determinations of the effective temperature require detailed modeling of the chemistry, the atmospheric structure and synthetic spectra (e.g., Burrows et al. 2001). Ideally, the spectral subtypes should correspond to a regular temperature scale, but this goal has not yet been reached for L and T dwarfs. The spectral classification schemes utilize the temperature dependent chemistry of the different elements but the appearance or disappearance of different atoms and molecules does not happen in evenly spaced temperature intervals (see chemistry below). So using the presence/absence of chemistry features alone for subtyping cannot lead to a temperature scale that linearly correlates with spectral subtype. Defining the subtypes so that they correspond to evenly spaced differences in effective temperatures requires that one knows the absolute temperature at the beginning and the end of the L dwarf sequence, and similarly at the start and end of the T dwarf sequence. This requires synthetic spectra to calibrate effective temperatures to the strength of certain absorption features so that practical flux ratios for subtyping can be defined.

The two classification schemes for L dwarfs by K99,K00 and M99 agree from late M to L3 but the subtype classification and the assigned $T_{eff}$ diverge significantly for later subclasses. K99 applied observed molecular and atomic features and chemical equilibrium calculations of Burrows & Sharp (1999) to estimate $T_{eff}$ ranges of 2000 to 1500 K between L0 to L8. M99 defined each L spectral subtype to be exactly 100 K wide and they used a $T_{eff}$ scale for a set of L dwarfs determined from optical lines of Cs I and Rb I by Basri et al. (2000).

These initially derived $T_{eff}$ probably need some revision because at the time, the applied chemistry of Burrows & Sharp (1999) used by K99 & K00, and the atmosphere models employed by Basri et al. (2000) used by M99 assumed chemical equilibrium without sedimentation of condensates and also excluded condensate opacities. However, since condensates are unavoidably present in L and T dwarfs they affect the emerging spectral flux, either by blocking and scattering radiation or by having removed gas opacity (see e.g., Tsuji et al. 1996a,b, 1999, Jones & Tsuji 1997, Ackerman & Marley 2001, Marley et al. 2002, Tsuji 2002).

Estimates for the effective temperatures in the spectral class L are about 2500-2000 K at M9/L0 for the coolest "true" stars, to about 1500-1200 K for the substellar objects near the L/T transition (Table 2). Within the T dwarf sequence a few objects have well constrained effective temperatures. For Gl 229B (T6) and Gl 570D (T8) these are 940±80 K and 805±20 K, respectively (Saumon et al. 2000; Geballe et al. 2001). The coolest T dwarf currently known is 2MASS J04151954-0935066 and estimates for its $T_{eff}$ are 600-750 K (Golimowski et al. 2004), and 760 K (Vrba et al. 2004). Overall, the uncertainties in the $T_{eff}$ scales of L and T dwarfs remain large (e.g., references in Table 2).

Table 2. Estimates of effective temperatures [a]

| M9/L0 [b] | L8-9/T0 | T8 | Source |
|---|---|---|---|
| 2500 | 1390 | 760 (T8/9) | Vrba et al. 2004 |
| 2400 | 1360 | -- | Dahn et al. 2002 |
| 2300 | 1450 | 600-750 (T9) | Golimowski et al. 2004 |
| 2250 | 1230 | | Burgasser et al. 2002a |
| 2200 | 1400 | -- | Stephens et al. 2001, Marley et al 2002 |
| 2200 | 1200 | -- | M99, Basri et al. 2000 |
| 2100 | 1300 | 800 (T8) | Geballe et al. 2001, 2002; Leggett et al. 2001, 2002 |
| 2030 | 1300 | 920 (T7.5) | Nakajima et al. 2004; Tsuji 2002 |
| 2000 | 1400 | -- | Schweitzer et al. 2001, 2002 |
| 2000 | 1300 | -- | K99, K00 |

[a] Typical uncertainties are ±100 K.
[b] Spectral type L according to K99

## 1.4. The case for condensate cloud layers

Condensate formation is important for understanding the atmospheric chemistry and spectra of low mass stars and substellar objects. Condensate cloud formation in an aging, cooling atmosphere seems unavoidable, hence condensate cloud layers are to be expected in L and T dwarfs. Cloud layer formation leads to drastic compositional and morphological atmospheric changes by removal of potential absorbers from the observable atmosphere and by blocking and scattering emerging spectral flux depending on how close they are to the photosphere. So in addition to chemical effects (see below), the physical effects of clouds (number and sizes of cloud layers, size of cloud particles) must be included in the atmospheric models that are needed to compute spectra. Physical cloud models have been studied by Lunine et al. 1989, Tsuji et al. 1996a,b, 1999, 2004 Jones & Tsuji 1997, Chabrier et al. 2000, Ackerman & Marley 2001, Allard et al. 2001, Marley et al. 2002, Tsuji 2002, Cooper et al. 2003, Tsuji & Nakajima 2003, Woitke & Helling 2003).

Condensation in a planetary, substellar or stellar atmosphere proceeds differently than condensate formation in a low gravity environment such as the solar nebula (or other protoplanetary disks) and stellar outflows (such as from giant stars). In a bound atmosphere, condensates precipitating from the gas at high temperatures ('primary condensates') settle due to the influence of gravity and form relatively thin discrete cloud layers. Consequently, the primary condensates cannot react with gas at higher altitudes above the condensate clouds, and are out of equilibrium with the overlaying atmosphere. Thus there are no secondary condensates from gas-solid or gas-liquid reactions at lower temperatures.

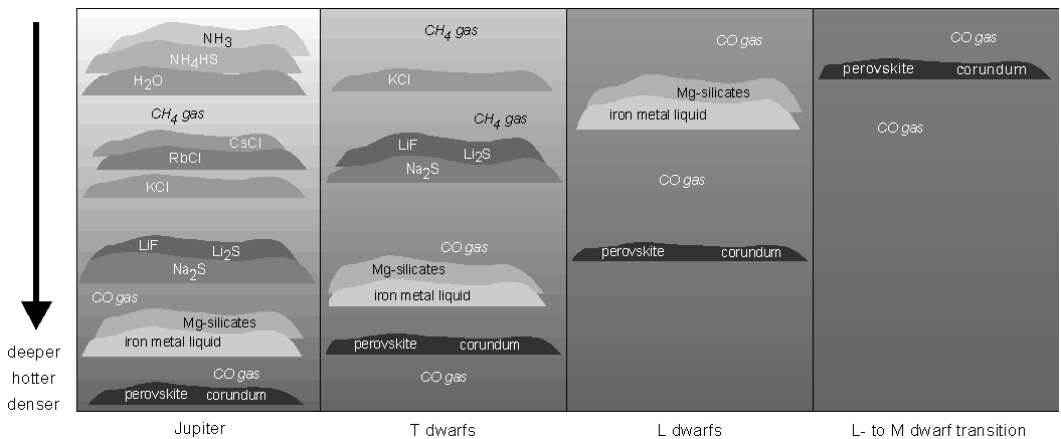

**Figure 1**. The change in cloud layer structure from cool to hot objects.

The cloud-layer condensation approach (occasionally called rainout) works well for the Jovian planets where refractory elements (e.g., Ca, Al, Mg, Si, Fe, Ti, V) must be sequestered into high-temperature clouds deep in the atmospheres (e.g., Lewis 1969, Barshay & Lewis 1978, Fegley & Prinn 1985a,b, 1986, Fegley & Lodders 1994). Hence it is plausible to apply this approach to the atmospheres of more massive, substellar objects (Fegley & Lodders 1996). Giant planets, T and L dwarfs show a wide range in effective temperatures and the number of cloud layers depends on the volatility of the different elements, so giant planets, T, and L dwarfs will have different numbers of cloud layers (Fig. 1). Refractory oxide clouds such as corundum clouds already appear in the photospheres of late M dwarfs (Jones & Tsuji 1997). Silicates and liquid iron metal can form cloud layers in the hottest L dwarfs.

On the other hand, Jupiter has water, ammonium hydrosulfide ($NH_4SH$), and solid ammonia as top-level clouds (Fig. 1). Deeper inside, alkali halide and sulfide clouds follow, then silicate and iron cloud layers. The deepest Jovian cloud layer consists of refractory Ca-Al- and Ca-Ti oxides. In the hotter T and L dwarfs, the cloud layer structure shown for Jupiter is successively stripped off at the top, and only the cloud layer with the most refractory condensates may be left in the hottest L dwarfs. Thus, looking at hotter substellar objects is like looking at deeper and deeper regions of Jupiter's atmosphere after removing the cooler atmospheric regions. The net effect of condensate cloud formation is that the atmosphere above a cloud is depleted in gases that contained the elements condensed into the clouds.

The following examples show how observable chemical tracers can be used to constrain the cloud layer model. The absence of silane ($SiH_4$) and the presence of germane ($GeH_4$) in the atmospheres of Jupiter and Saturn is due to depletion of refractory Si, but not of volatile Ge, by magnesium silicate condensate cloud formation deep in their atmospheres (Fegley & Lodders 1994). Silicon is much more abundant than Ge in a solar composition gas (atomic Si/Ge ~8300) but $SiH_4$ is not observed on either Jupiter or Saturn (observational upper limits are $SiH_4/H_2$ ~ $1\times10^{-9}$ by volume (1 ppbv)). For comparison, the solar $Si/H_2$ molar ratio is $7.09\times10^{-5}$, which is about 71,000 times larger than the observational upper limit on the silane abundance. In contrast, $GeH_4$ is observed with a $GeH_4/H_2$ ratio ~0.7 ppbv on Jupiter and ~0.4 ppbv on Saturn. These values are less than the solar $Ge/H_2$ molar ratio of 8.5 ppbv because not all Ge in the atmospheres of Jupiter and Saturn is present as $GeH_4$ (Fegley & Lodders 1994). The presence of silicate clouds in T dwarfs is testable with searches for $SiH_4$ in T dwarfs and Si, SiO, SiS or SiH in L dwarfs. In the atmosphere above the $Mg_2SiO_4$ and $MgSiO_3$ clouds, Si-bearing gases should be highly depleted and be absent at the highest altitudes.

The *Galileo* entry probe mass spectrometer (GPMS) detected $H_2S$ at about 3 times the solar $S/H_2$ ratio in Jupiter's atmosphere (Niemann *et al.* 1998). This is consistent with depletion of Fe-metal by Fe-cloud condensation (e.g., Lewis 1969, Barshay & Lewis 1978, Fegley & Lodders 1994). If Fe cloud formation did not occur, $H_2S$ would be completely absent from the atmosphere of Jupiter because formation of FeS (troilite) from Fe metal grains with $H_2S$ gas at ~700 K consumes all $H_2S$ gas (solar Fe/S~2). Hence the Fe-metal condensation cloud layer model is in accord with the GPMS observations of $H_2S$ on Jupiter.

Monatomic K gas is present in the atmospheres of the T dwarfs Gl 229B and Gl 570D (e.g., Burrows et al. 2000, Geballe et al. 2001). The observed K abundances require that refractory rock-forming elements such as Al, Ca, and Si be removed by condensate cloud formation deep in the atmosphere of Gl 570D and Gl 229B (e.g., Geballe et al. 2001). Otherwise, potassium would condense into silicate minerals such as $KAlSi_3O_8$ (orthoclase) at high temperatures, and K gas would be depleted from the observable atmosphere (Lodders 1999, Burrows et al. 2000).

## 1.5. Chemistry of selected elements

This section summarizes the chemistry of C, N, O, the major rock-forming elements Ca, Al, Mg, Si, Fe, Cr, and of Li and the other alkali elements (Na, K, Rb, Cs) that significantly influence the spectra of low mass objects. The chemistry is derived from thermochemical equilibrium computations, and where necessary kinetic considerations (see Fegley & Lodders 1994, Lodders 1999, 2002, Lodders & Fegley 2002). The atmospheric chemistry for Jupiter and Saturn of essentially all naturally occurring elements is discussed by Fegley & Lodders (1994) and is a useful guide for the chemistry of other elements in the coolest low mass objects. However, some differences may arise because Jupiter and Saturn are enriched in elements heavier than He relative to solar abundances.

Relatively recently the solar abundances of C and O were significantly revised downward and abundances of many other elements were updated (see Lodders 2003). The solar abundances are routinely used to model objects of solar metallicity hence the revisions in solar abundances change previous results for the temperature and pressure dependent distribution of the elements between gases and condensates. However, compared to previous results the new solar abundances do not change the types of gases and condensates that are present in substellar objects and only cause relative shifts of species abundances as functions of temperature and pressure. Here all results from thermochemistry are shown for the solar system abundances in Lodders (2003).

The major chemistry features as a function of total pressure ($P_{tot}$) and temperature are summarized in Figs. 2a,b and are described in the following subsections. The Figures also show model atmospheres for an M dwarf ($T_{eff}$ = 2200 K, dust-free; Tsuji et al. 1996a), Gl 229B ($T_{eff}$ = 960 K, Marley et al. 1996), and Jupiter to give some orientation which chemical features are relevant for objects of different mass.

### 1.5.1. Carbon, nitrogen, and oxygen chemistry

After hydrogen, C, N, and O are the next most abundant chemically reactive elements. The review of the C, N, and O chemistry is kept brief because a detailed description is in Lodders & Fegley (2002). Much of the chemistry and therefore the spectral appearance of substellar objects is governed by the distribution of C and O between water, methane, and carbon monoxide according to the net thermochemical reaction

$$CO + 3 H_2 = CH_4 + H_2O$$

The distribution of C and O between these gases also determines the stability of many other gases and condensates containing other elements, which makes this equilibrium of fundamental importance for understanding the chemistry of substellar objects.

Figures 2a and 2b illustrate carbon equilibrium chemistry as a function of P and T. The dotted curve labeled CO = $CH_4$ is the equal abundance curve for these two gases. Carbon monoxide is more abundant than $CH_4$ in the lower pressure and higher temperature region to the left of the curve (the CO field). Methane is more abundant than CO in the higher pressure and lower temperature region to the right of the curve (the $CH_4$ field). The shift from CO as the major C-bearing gas to $CH_4$ as the major C-bearing gas is gradual and both gases are always present, even if one is overwhelmingly dominant. This point is illustrated by Figure 2 in Fegley and Prinn (1989) which shows logarithmic CO/$CH_4$ contours of … 0.01, 0.1, 1, 10, 100 …as a function of temperature and $P_{tot}$ for solar elemental abundances of H:C:O = 1446:1:0.6.

Late M dwarfs with relatively high temperatures and low total pressures in their outer atmospheres plot in the CO field (Fig. 2a,b). In their observable atmospheres essentially all C is bound to CO and most of the rest of oxygen not in CO is bound to water. On the other hand, Jupiter's atmospheric P-T profile falls into the methane field and only a very small amount of CO is present (but also see kinetics below) and essentially all C is found as methane and O as water.

The P-T profile of the relatively late T dwarf Gl229 B (T6) with an effective temperature of 970 K falls into the methane field but crosses into the CO field at about 1600 K. Thus, CO is an abundant gas in the deeper atmosphere of Gl229B and convective mixing of kinetically stable CO from the deeper interior into the observable atmosphere can occur. For example Noll et al. (1997) detected the fundamental band of CO (4.7 μm) in Gl 229B, revealing a CO/$CH_4$ ratio ~$10^3$ times larger than expected from chemical equilibrium (see Fegley & Lodders 1996, Griffith & Yelle 1999, Saumon et al. 2000, 2003, and kinetics below).

The L/T dwarfs transition is determined by the appearance of abundant methane and the CO = $CH_4$ boundary can provide an estimate for the effective temperature at the L/T transition. At the total pressures characteristic of substellar dwarf atmospheres methane is only more abundant than CO at temperatures below 1500-1200 K (Fig. 2b). This temperature range is in accord with temperature estimates at the L/T transitions listed in Table 2. However, the kinetics of CO destruction (see below) is also important in late L and early T dwarfs because it affects methane abundances which in turn affect the onset of methane bands. It also affects the atmospheric structure because methane is the second strongest absorber after $H_2O$ in T dwarfs (e.g., Saumon et al. 2003).

Figure 2 also shows the curve of equal gas abundances for molecular nitrogen and ammonia ($N_2$=$NH_3$, dotted curve). This is shifted to higher $P_{tot}$ and lower temperatures than the CO=$CH_4$ boundary. This makes the $N_2$/$NH_3$ a promising temperature probe for the coolest T dwarfs.

Ammonia was first detected in the H and K bands of the near-IR spectrum of the T6 dwarf Gl229B (Saumon et al. 2000). The P-T profile of Gl229B stills falls into the field where $N_2$ is more abundant than $NH_3$ but approaches the $NH_3=N_2$ boundary at lower T where $NH_3$ abundances may become large enough for detection. The models by Burrows et al. (2003) predict observable amounts of $NH_3$ at 1.5, 1.95 μm (H band) and 2.95 μm (K-band) for substellar dwarfs with effective temperatures below 600 K. These expectations are consistent with the observations by Knapp et al. (2004) and Golimowski et al. (2004) that $NH_3$ absorptions are absent in the H and K bands for the coolest known T dwarf 2MASS J04151954-0935066 with $T_{eff}$ ~700 K.

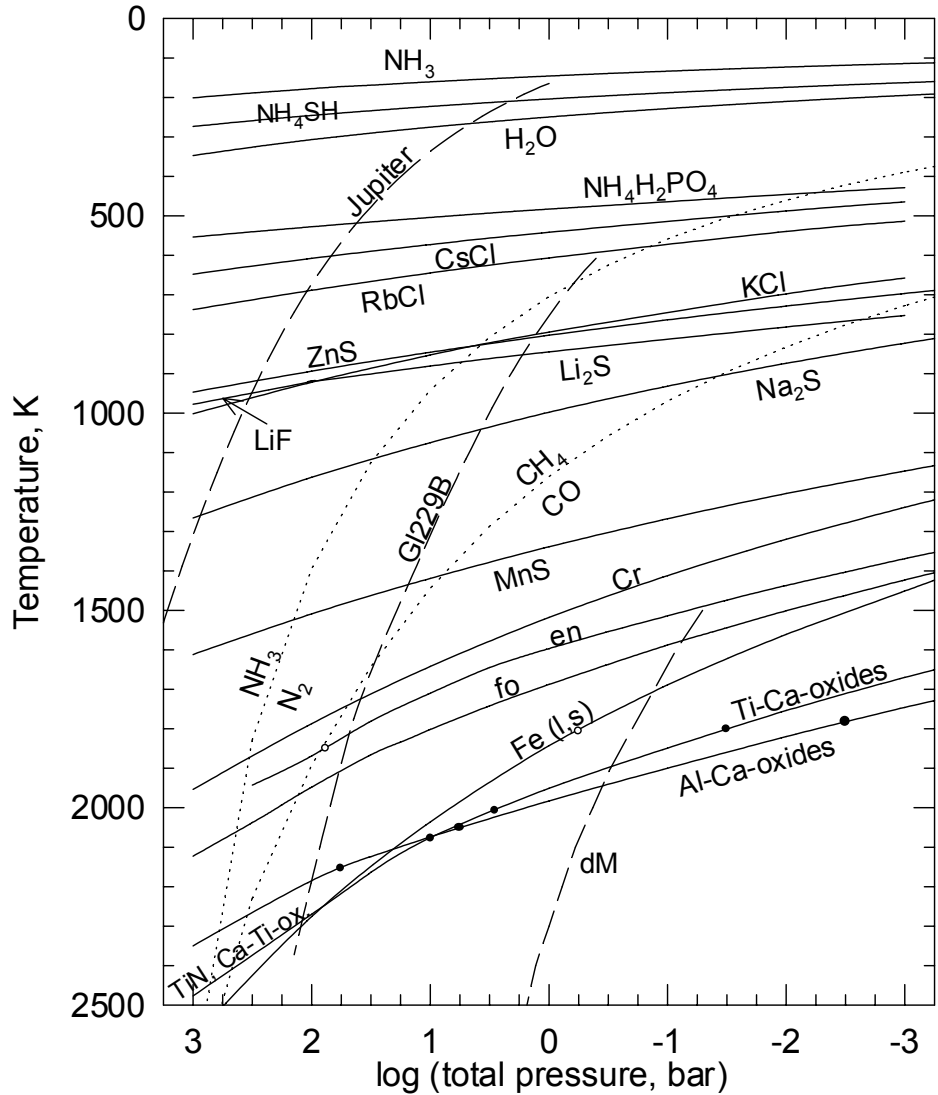

**Figure 2a**. Major features of the equilibrium chemistry for a solar gas as a function of temperature and total pressure. The condensation temperatures of major element and alkali element condensates are the solid lines. The constituent condensate elements are removed from the gas above these curves. The curve labeled "Ca-Al-oxide" shows which Al-bearing condensate forms first as a function of total pressure. At low pressures, corundum forms. With increasing pressure hibonite, grossite, and gehlenite are the first Al-bearing condensates and the transitions are indicated by the black dots on the Ca-Al-oxide curve. Similarly, the curve "Ca-Ti-oxide" shows that at low total pressure, perovskite is the first Ti bearing condensate, then $Ca_4Ti_3O_{10}$, then $Ca_3Ti_2O_7$, and at the highest $P_{tot}$, TiN. The white dots on the enstatite and iron condensation curves indicate melting points. The long-dashed curves indicate atmospheric P-T conditions for Jupiter, Gl229 B ($T_{eff}$ = 960 K, Marley et al. 1996), and a M-dwarf ($T_{eff}$ = 2200 K, dust-free; Tsuji et al. 1996a). The dotted curves show equal gas abundances for the pairs $CO/CH_4$, and $N_2/NH_3$.

Ammonia absorption in T dwarfs is best detectable at mid-infrared wavelengths, and $NH_3$ absorptions at 7.8 and 10 μm were recently reported by Roellig et al. (2004) in spectra for the composite T1/T6 binary system ε Ind (Ba/Bb) with the Spitzer Space Telescope. Note that one of the binary components has essentially the same spectral type as Gl229B so that the presence of $NH_3$ in T6 dwarfs seems certain. The mid-IR absorption features of $NH_3$ seem to be absent in the spectrum of a L8 dwarf (Roellig et al. 2004). Given that the L/T transition is marked by the onset of methane bands in the near IR and that there are considerable temperature differences between the $CO=CH_4$ and $N_2=NH_3$ boundaries, large detectable amounts of ammonia are unlikely near the L/T transition. Instead, the appearance of ammonia in the near- and mid infrared bands will probably become a marker for the T/Y dwarf transition.

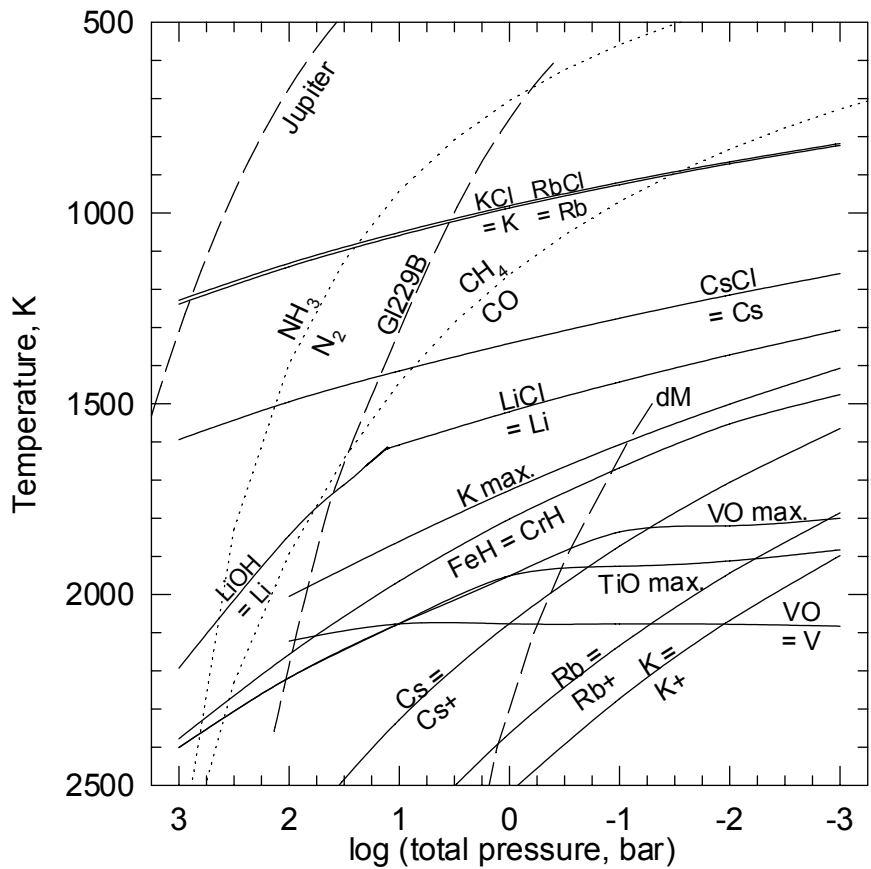

**Figure 2b**. Key features of gas phase equilibrium chemistry for a solar gas. Curves of equal abundances for the pairs $CO/CH_4$ and $N_2/NH_3$ are shown by dotted curves and for several other pairs of gases as solid curves. Also shown are curves where TiO, VO and K reach their maximum in abundances (e.g., TiO max.) The long-dashed curves indicate atmospheric P-T conditions for Jupiter, Gl 229B, and an M-dwarf (see Fig. 2a).

However, kinetics (see below) are also important for the $N_2$ to $NH_3$ conversion. For example, ammonia was found to be 50% depleted in the K band spectrum of Gl 229B which suggests that $N_2$-rich gas is transported upward in the atmosphere (Saumon et al. 2000). Observations of $NH_3$ in mid-infrared spectra (such as reported by Roellig et al. 2004) probe further levels of the atmospheres. Quantitative analyses of $NH_3$ bands at different atmospheric levels then should provide useful information on the mixing processes in substellar dwarf over several pressure scale heights (Saumon et al. 2000, 2003).

The larger abundances of O and N lead to massive cloud layers at the lowest temperatures when water, $NH_4SH$, and solid $NH_3$ condense (Fig. 2a) which effectively remove all N and O from the atmosphere above the clouds. However, these clouds are not expected within the atmosphere of the latest T dwarfs. As mentioned earlier, objects with water clouds will be members of a new spectral class (Y dwarfs) or larger planetary objects.

**1.5.2. Refractory rock-forming elements: Al, Ca, Ti and V**

The major rock forming elements (Al, Ca, Ti) can form high temperature condensate cloud layers. Although not a major element, V is on the list of elements here because it is observed in dwarfs and its chemistry is similar and related to that of Ti. The condensation temperatures of Ca-Al-oxides and Ca-Ti-oxides are shown in Fig. 2a by the solid lines. Curves for equal abundances of certain pairs of gases and curves for maximum abundances of a given gas are shown in Fig. 2b.

At temperatures below 2500 K but above temperatures where the first condensates appear (Fig. 2a,b), the major gases are monatomic Ca; Al, AlH, $Al_2O$, AlF, and AlCl; TiO; V and VO (at $P_{tot}=1$ bar). The high temperature distribution of vanadium is of interest because the V/VO abundance ratio could serve as a good temperature indicator in M-dwarfs. High temperatures favor V as the dominant gas and low temperatures VO, and the boundary where V and VO reach equal abundances (at T ~2080) is relatively insensitive to $P_{tot}$ (Fig. 2b).

The more pressure sensitive transition from Ti to TiO occurs at temperatures above 2500 K for the total pressures shown here. Once VO is the major V-bearing gas, $VO_2$ gas increases in abundance with decreasing temperature, and similarly, $TiO_2$ gas gains at the expense of TiO gas. The curves where TiO and VO gas abundances reach their maximum are shown in Fig. 2b. At $P_{tot}$ >10 bar, VO and TiO remain the dominant gases until all V and Ti gases are removed into condensates. This is the reason why the maximum TiO and VO abundances coincide with the condensation temperatures of the Ca-Ti-oxides.

When a condensate appears, its constituent elements are removed from the gas of the overlaying, cooler atmosphere above the condensate cloud. The abundances of *all* gases that contain the constituent element(s) of the condensate drop, e.g., once Ti condenses, the abundances of gaseous TiO, Ti, $TiO_2$, etc. are all reduced.

The most refractory major condensates contain Al-, Ca-, and Ti (Fig. 2a) and the specific mineralogy of the initial condensate depends on total pressure (Lodders 2002). The first Ca-Ti-bearing condensate as a function of $P_{tot}$ are perovskite ($CaTiO_3$) below 0.03 bar, $Ca_4Ti_3O_{10}$ from 0.03 to 3.2 bar, $Ca_3Ti_2O_7$ from 3.2 to ~10 bars, and above ~10 bars, TiN (osbornite) replaces the Ca-Ti-oxides as the first Ti-bearing condensate. The P-T range at which a certain Ca-Ti-oxide is stable is indicated by black dots on the "Ca-Ti-oxide" condensation curve in Fig. 2a. Vanadium condenses into solid solution with Ca-Ti-oxides and can also form a solid solution with TiN at the highest total pressures.

The gradual disappearance of TiO and VO bands from the spectra of early L dwarfs is most plausibly explained by condensate formation. Once the refractory elements Ti and V are removed into cloud layers, they are no longer available to form gaseous TiO and VO (Fegley & Lodders 1994, 1996, Lodders 2002).

The mineralogy of the first Ca-Al-bearing condensates is also a function of total pressure (Lodders 2002) At the lowest $P_{tot}$ ($<3 \times 10^{-3}$ bar) corundum ($Al_2O_3$) is the initial condensate. With increasing $P_{tot}$, the initial condensates are hibonite ($CaAl_{12}O_{19}$) from $3 \times 10^{-3}$ to 5.6 bars, grossite ($CaAl_4O_7$) from 5.6 to ~63 bar, and melilite (a solid solution of gehlenite and akermanite) above 63 bar. The condensation temperatures of the Ca-Al-oxides are similar to those of the Ca-titanates and formation of calcium aluminate clouds leads to the disappearance of the Ca I and Al I lines and CaH bands near the M to L dwarf transition (Table 1).

### 1.5.3. Abundant condensates from rock-forming elements: Fe, Cr, Mg, Si

The solar abundances of Fe, Mg and Si are much larger than those of the refractory rock-forming elements (Ca, Al, Ti) and condensates of Mg, Si and Fe form the major mass of high temperature condensates. Chromium is included in the list here because its chemistry is comparable to that of Fe so they are described together.

After condensation of the refractory elements Ca, Al, and Ti, iron metal forms the next cloud layer (at 1843 K at 1 bar). Before condensation occurs, monatomic Fe is the major Fe-bearing gas in dwarf atmospheres. Chromium behaves similarly – until condensation of Cr-metal (1518 K at 1 bar) the major Cr-bearing gas is monatomic Cr (note that condensation of $Cr_2O_3$ instead of Cr metal only occurs at lower $P_{tot}$ ~$10^{-3}$ bar).

The hydrides FeH and CrH are always less abundant than the respective monatomic gases and the hydride gas abundances actually decrease with decreasing temperature (at constant $P_{tot}$) according to the equilibrium M (g) + 0.5 $H_2$ = MH (g) where M = Fe or Cr. We note that other metal hydrides such as NiH and CoH also appear in a similar manner.

Although less abundant than the monatomic gases, the hydrides of Fe and Cr strongly shape the spectra of L and T dwarfs. The major reason for increasingly stronger FeH and CrH bands within the L sequence up to spectral types L4-L6 is that other absorbers such as TiO and VO gradually disappear into condensates (see previous section) so that absorption features from other compounds become more apparent. The decline in FeH band strength starting with spectral type L5 must be related to the condensation of Fe metal which reduces the abundances of all Fe-bearing gases (i.e., monatomic Fe and FeH). Condensation of Cr metal reduces the abundances of monatomic Cr and CrH and occurs at lower temperatures than for Fe (Fig. 2a) which explains the decline of CrH band strength at a later spectral type (L6). After Fe metal condensation, abundances of FeH drop and at some point CrH and FeH abundances are equal. This is shown by the curve CrH=FeH in Fig. 2b. If spectroscopic properties of CrH and FeH are sufficiently similar, this curve approximates these equal abundances to equal band strengths, which is observed around spectral type L4 (Table 1).

At temperatures between Fe and Cr metal cloud condensation, a substantial cloud layer mass develops when Mg and Si condense as forsterite ($Mg_2SiO_4$) and enstatite ($MgSiO_3$) within a relatively narrow temperature interval (1688 and 1597 K, resp. at 1 bar). This removes opacity sources associated with the major Mg and Si-bearing gases (Mg, Si, SiO, SiS, SiH) from the overlaying gas.

A comparison of the condensation temperatures of the Mg-silicates and those of Fe and Cr and the behavior of the FeH and CrH bands with spectral type suggests that condensation of the Mg-silicate cloud layer should leave its spectral signature in the mid-L sequence (~L4-L6). So far, no direct chemical tracers have been observed that can confirm the presence of Mg-silicate clouds. These clouds may be traceable by the strength of neutral monatomic Mg and Si lines, or in the case of Si, observations of band strengths of SiO, SiS and various Si hydrides (depending on $P_{tot}$). However, the latter require near to mid IR observations and bands of the Si-bearing molecules could be swamped by those other strong absorbers (e.g., FeH, CrH, CO, $H_2O$) still present when the silicates form. The Mg I lines may be more prominent that Si I lines because Mg I is the dominant Mg gas whereas Si I gas is a minor fraction of all Si-bearing gases. Very weak Mg I lines can be seen in the J-band spectra of late M dwarfs and are probably present in L2 dwarfs (Mc03) but these weak lines are not useful for monitoring any systematic decrease of Mg through the L-sequence.

Any clouds close to or above the photosphere will leave their mark on infrared colors. The effect of clouds is also important for understanding the weakening of some spectral features in the mid-L sequences and the strengthening again of the same features (FeH, K I, see below) in the early T sequence. Detailed models suggest that the clouds are expected to be most optically thick around L4-L6 (see, Ackerman & Marley, 2001, Burrows et al. 2001, Tsuji 2002, Knapp et al. 2004).

Only in cooler objects where the silicate clouds sit deeper below the photosphere does the observable part of the atmosphere become more transparent again.

Figure 2 shows that Fe, forsterite, and enstatite may condense as liquids at sufficiently high pressures. For example, Fe liquid forms at T $\geq$1809 K, forsterite liquid at T $\geq$2163 K, and enstatite liquid at T $\geq$1851 K. Many models of dwarf atmospheres now include dust opacities but mainly of solid dust. However, for several objects the derived P-T conditions may fall into the stability fields of liquid condensates. The formation of dust particles or suspended liquid droplets (hazes or aerosols) can have different effects on atmospheric opacities and needs to be considered in atmospheric modeling. Liquid condensates may also dissolve other elements, have complex compositions, and form at lower temperatures.

### 1.5.4. The alkali elements Na, K, Rb, and Cs

Atomic alkali element lines are prominent in all L dwarfs and remain detectable in the hottest T dwarfs. With decreasing temperature, more of the alkali atoms convert into halide gases and also condense into sulfide and/or halide clouds. As before, we follow the chemistry from high to low temperatures in Fig. 2a,b and use 1 bar as reference total pressure for the discussion.

Monatomic Na and K are the major Na and K gases at the highest temperatures relevant for L and T dwarfs; in M dwarfs $K^+$ is important in addition to neutral K (Fig. 2b). Monatomic Na remains the dominant Na gas until $Na_2S$ condensation at ~1000 K but the abundances of other Na-bearing gases such as NaCl steadily increase with decreasing temperature. In contrast, KCl becomes the dominant K-bearing gas at temperatures below the boundary at which monatomic K and KCl abundances are equal (K = KCl; ~980 K at 1 bar, see Fig. 2b). Formation of a KCl condensate cloud layer at ~800 K leads to removal of all K-bearing gases from the overlaying atmosphere. Thus, monatomic Na is mainly depleted through condensation of $Na_2S$ and monatomic K is depleted by conversion into KCl gas. These chemical changes are expected for temperatures below 1000 K (at 1 bar).

The strength of the Na I doublet (0.5890 & 0.5896 μm) begins to decline and to broaden at L1 and is barely perceptible at L8. The spectral region of the K I resonant doublet lines (0.7665 & 0.7699 μm) develops into a broad depression throughout the L sequence (K99). This behavior is due less to the chemical changes because above 1200 K, the lowest estimate for $T_{eff}$ at the L/T transition (Table 2), monatomic Na and K are still the most abundant Na and K-bearing gases, although their chlorides become more abundant as temperature decreases. Instead, pressure-broadening of the Na I and K I resonance doublets diminishes the Na and K lines strengths.

Rubidium and Cs have the lowest ionization potentials of naturally occurring elements so $Rb^+$ and $Cs^+$ ions are the dominant gases at the highest temperatures shown in Fig. 2b. The abundances of $Rb^+$ and $Cs^+$ decrease with decreasing temperature and the curves labeled $Rb^+$ = Rb and $Cs^+$ = Cs indicate where the ion and neutral atom abundances are equal. At intermediate temperatures monatomic

Rb and Cs are the most abundant Rb and Cs gases. RbCl and CsCl gas become more abundant at lower temperatures. The equimolar curves (Rb = RbCl and Cs = CsCl) are reached at 987 and 1342 K, respectively (at 1 bar; Fig. 2b). Further drops in temperature then allow condensation of solid RbCl (607 K at 1 bar) and CsCl (542 K at 1 bar) which depletes the atmosphere in gaseous RbCl and CsCl.

**1.5.5. Lithium chemistry**

Lithium is of particular interest because low mass dwarfs (<0.065 $M_{sun}$) are not expected to burn Li and their protosolar abundances of Li should be preserved. In that case, the substellar nature of a low-mass object could be confirmed by the observation of monatomic Li (Rebolo et al. 1992). However, this Li-test is of limited use in cooler dwarfs because the abundance of atomic Li (gas) is reduced below the bulk Li abundance when other Li-bearing gases such as LiOH and LiCl form and consume monatomic Li.

Lithium chemistry is more sensitive to $P_{tot}$ than the chemistry of the other alkali elements, and several other gases (LiOH, LiH, LiF) gases are more important in addition to LiCl (Lodders 1999). Monatomic Li is the dominant Li-bearing gas down to 1520 K (at 1 bar) where LiCl becomes the major Li-bearing gas (Fig. 2b). At $P_{tot}$ above ~30 bar, monatomic Li converts to LiOH which is then the most abundant Li gas. The Li=LiOH boundary (restricted to high $P_{tot}$) is much more sensitive to total pressure than the Li=LiCl boundary at lower $P_{tot}$. Additional Li-bearing gases (not shown) become abundant at lower temperatures until condensation of $Li_2S$ (at low $P_{tot}$) and LiF (at high $P_{tot}$) occurs (Fig. 2a.)

Lithium chemistry is influenced by the position of the CO = $CH_4$ boundary and LiOH is generally more important when $CH_4$ is more abundant than CO, and LiCl is generally more important when CO is more abundant than methane. The reason for this is that more $H_2O$ is available when $CH_4$ forms via: $CO + 3 H_2 = CH_4 + H_2O$ so that the reaction $LiCl + H_2O = LiOH + HCl$ also proceeds.

The increasing formation of Li-bearing molecules at lower temperatures places limits on the use of monatomic Li as the sole measurement of the Li content in a cool atmosphere. In order to obtain a measurement of the bulk Li abundance in L dwarfs the abundance of LiCl needs to be measured in addition to monatomic Li. Recently Weck et al. (2004) determined the spectroscopic properties of LiCl gas, which, in principle, would allow such determinations. However, as noted by Weck et al. (2004), suitable bands of LiCl are only in the mid infrared (15.8 μm) and these are most likely also blended by water absorptions.

**1.5.6. Notes on a few other element condensates**

For reference, we plotted some condensates of other, more abundant elements (Mn, Zn, P, S) expected in substellar dwarf atmospheres. Manganese sulfide (MnS) condenses between Cr-metal and $Na_2S$, and is expected to form a small cloud layer in late L-dwarfs (at 1340 K at 1 bar). Another sulfide, ZnS condenses at 800 K (at 1 bar), at similar temperatures as KCl in dwarfs around the L/T transition. A low temperature condensate that may appear already in very late T dwarfs is $NH_4H_2PO_4$ and is certainly expected for Y dwarfs. Another very low temperature condensate, $NH_4SH$, should also appear as a larger cloud layer in Y dwarfs since this condensate removes most of the abundant $H_2S$ from the atmosphere. Details for the chemistry of these elements are not yet described for substellar dwarf atmospheres, but their chemistry in Jupiter and Saturn is described in Fegley & Lodders (1994).

## 1.6. Kinetics

Thermochemical reactions in the atmospheres of planets and low mass substellar objects take place in dynamic environments with only limited time available for reactions to reach chemical equilibrium. Furthermore, different chemical reactions take place at different rates, which generally decrease exponentially with decreasing temperature.

Since the mid 1970s, telescopic observations showed CO, $PH_3$, $GeH_4$, and $AsH_3$ on Jupiter and Saturn at abundances much greater than expected from chemical equilibrium considerations. Vertical convective mixing brings these species to observable regions of Jupiter's and Saturn's atmospheres from deeper levels where their chemical equilibrium abundances are orders of magnitude larger (Fegley & Lodders 1994). Similar effects may be important for CO, $N_2$, $PH_3$, $GeH_4$, $AsH_3$, and HCN in the atmospheres of sub-stellar objects (Fegley & Lodders 1996). We briefly review the theory underlying thermochemical disequilibrium chemistry in planetary and sub-stellar atmospheres and then describe possible consequences for chemistry of C, N, Cs, and Fe.

Carbon monoxide is a good example to explain the basic theory. As mentioned earlier, CO is the dominant carbon gas at high temperatures and low pressures, while $CH_4$ is the dominant carbon gas at low temperatures and high pressures. The CO equilibrium abundance decreases strongly with decreasing temperature. The altitude range over which [CO], the CO number density, decreases by a factor of $e$ is the CO chemical scale height $h_{chem}$

$$h_{chem} = -\frac{[CO]/t_{chem}}{\dfrac{d}{dz}\left([CO]/t_{chem}\right)}$$

The $t_{chem}$ in this equation is the chemical lifetime for CO destruction. It is defined by

$$t_{chem} = -\frac{[CO]}{d[CO]/dt}$$

The conversion between CO and $CH_4$ occurs via the net thermochemical reaction

CO + 3 H$_2$ = CH$_4$ + H$_2$O

However, three H$_2$ molecules do not simultaneously collide with a CO molecule to yield one CH$_4$ and one H$_2$O molecule. Instead, the net reaction plausibly proceeds via a series of elementary reactions (i.e., the actual chemical steps that take place) such as that proposed by Prinn & Barshay (1977):

H$_2$ = H + H
CO + H$_2$ = H$_2$CO (formaldehyde)
H$_2$CO + H$_2$ → CH$_3$ + OH
CH$_3$ + H + M = CH$_4$ + M
<u>H + OH + M = H$_2$O + M</u>
CO + 3H$_2$ = CH$_4$ + H$_2$O        Net Reaction

Here the M that appears in some of the equations is any third body, statistically H$_2$ or He in the atmosphere of a sub-stellar mass object, and it is necessary to absorb the energy released by forming the H$_3$C-H bond in methane and the H-OH bond in water. The = sign denotes reactions that are in equilibrium in the hot, deep atmospheres of gas giant planets and sub-stellar mass objects. The direction of a reaction is indicated by the arrow. This series of elementary reactions sums to CO + 3 H$_2$ = CH$_4$ + H$_2$O, but the actual reaction sequence is more complicated than suggested by that one reaction.

The chemical lifetime $t_{chem}$ for CO destruction is computed from the rate of the slowest elementary reaction (the rate determining step), which is the reaction

H$_2$CO + H$_2$ → CH$_3$ + OH

The t$_{chem}$ for CO destruction is

$$t_{chem} = \frac{[CO]}{[H_2CO][H_2]k}$$

The rate constant $k$ for the rate-determining step is

$$k = 2.3 \times 10^{-10} \exp(-36,200/T) \text{ cm}^3 \text{ s}^{-1}$$

The $t_{chem}$ and $h_{chem}$ for CO decrease dramatically with decreasing temperature in atmospheric regions where the activation energy factor ($E_a/R = 36,200$) is much larger than temperature $T$. Carbon monoxide is destroyed by conversion to CH$_4$ + H$_2$O in gas parcels transported upward by convective mixing. The convective mixing time t$_{mix}$ is

$$t_{mix} \sim H^2 / K_{eddy}$$

The *H* and *K*$_{eddy}$ in this equation are the pressure scale height and the vertical eddy diffusion coefficient. The pressure scale height is a function of atmospheric temperature, mean molecular weight ($\mu$), and gravity

$$H = \frac{RT}{\mu g}$$

The vertical eddy diffusion coefficient is estimated from the convective heat flux and is typically $10^7 - 10^9$ cm$^2$ s$^{-1}$ for the tropospheres of Jupiter and Saturn. Destruction of CO stops at the level where

$$t_{chem}(\text{CO}) = t_{mix} \sim H^2 / K_{eddy}$$

$$h_{chem}^2 \ll H^2$$

This level is the quench level. For example, Prinn & Barshay (1977) calculated a quench level of 1100 K for CO on Jupiter where $h_{chem}$ (CO) ~25 km versus *H* ~220 km.

Several other molecules such as HCN and N$_2$ are analogous to CO because they also have bonds with large dissociation energies (CO 11.09 eV, N$_2$ 9.76 eV, CN 7.75 eV). The activation energy factors for destruction of N$_2$ ($E_a/R$ = 81,515) and HCN ($E_a/R$ = 70,456) are also much larger than *T* in the atmospheres of Jovian planets and sub-stellar mass objects. Thus, HCN and N$_2$ are also expected to be quenched at high temperature with the exact temperature depending upon the considerations above.

In contrast, several other species such as monatomic Cs and FeH are unquenchable in the atmospheres of gas giant planets and sub-stellar mass objects because the activation energy factors are effectively zero. For example, with decreasing temperature Cs is probably converted into CsCl by the elementary reaction

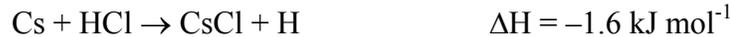
Cs + HCl → CsCl + H          $\Delta H = -1.6$ kJ mol$^{-1}$

Clay & Hussain (1990) measured a rate constant $k = 5.1 \times 10^{-12}$ cm$^3$ s$^{-1}$ at 828 K. The estimated temperature dependence of the rate constant is

$$k = 5 \times 10^{-10} \exp(-3800/T) \text{ cm}^3 \text{ s}^{-1}$$

This estimate is based on the measured value of Clay and Hussain (1990) and the temperature dependences for the rate constants for the analogous conversions of Li to LiCl, Na to NaCl, and K to KCl. The chemical lifetime of Cs is

$$t_{chem}(\text{Cs}) = (k[\text{HCl}])^{-1}$$

For example, calculations for an otherwise solar composition gas with a metallicity of –0.5 dex give $t_{chem}$ (Cs) values ≤ 1 second at 1 bar for T ≥ 1100 K. This shows that monatomic alkali to alkali halide conversions occur on very fast time scales that are much faster than plausible mixing timescales (such as on Jupiter). Griffith & Yelle (2000) also concluded that the observation of Cs in Gl229B, where CsCl instead of Cs is expected at the low temperatures, cannot be caused by the presence of Cs from convection as initially proposed by Burrows & Sharp (1999). Instead it is more likely that the observations of Cs probe a deeper, hotter atmospheric level where Cs is the dominant Cs gas (see Fig. 2b).

No kinetic data are available for destruction of FeH gas. However, several plausible destruction reactions are highly exothermic at 1000 K and should have activation energy factors close to zero

$$FeH + OH \rightarrow H_2O + Fe \quad \Delta H = -344 \text{ kJ mol}^{-1}$$
$$FeH + CH_3 \rightarrow CH_4 + Fe \quad \Delta H = -288 \text{ kJ mol}^{-1}$$
$$FeH + H \rightarrow H_2 + Fe \quad \Delta H = -282 \text{ kJ mol}^{-1}$$
$$FeH + HS \rightarrow H_2S + Fe \quad \Delta H = -228 \text{ kJ mol}^{-1}$$
$$FeH + FeH \rightarrow H_2 + 2Fe \quad \Delta H = -120 \text{ kJ mol}^{-1}$$

These reactions may proceed at rates close to the gas kinetic rate ($10^{-10} - 10^{-9}$ cm$^3$ s$^{-1}$) and have chemical lifetimes much less than the vertical mixing times in the atmospheres of sub-stellar mass objects.

Thus, the observations of strengthening FeH bands within the T dwarf sequence cannot be related to convective upward mixing of FeH gas as favored by Nakajima et al. (2004). This possibility was already considered by Burgasser et al. (2002b) who concluded that the fragile molecule FeH (and, similarly CrH with low dissociation energies 1.63 eV and 1.93 eV, respectively) precludes quenching of the destruction reactions of FeH and CrH from high temperatures.

## 1.7. Summary: A chemical temperature scale

Chemistry can be used to develop a 'chemical temperature scale', which should help to constrain effective temperatures within the L and T dwarf sequences. A chemical temperature scale is model-atmosphere independent and only uses the appearance (e.g., molecular gas formation) and disappearance (e.g., removal of gases by condensation) of certain compounds. However, the chemical temperature scale cannot easily be related to an effective temperature scale because observations at different spectral wavelength track chemical changes that occur at different depth (and thus temperatures) in the atmospheres. Furthermore, the temperatures derived from chemistry must be lower limits and are probably 100-200K less than the effective temperatures because the chemistry observed through the absorption features occurs within or above the photosphere. Keeping these limits in mind, we use the condensation temperatures and curves for gas abundance from Figs. 2a,b to relate the chemistry to spectral types. The major chemical features from Figs. 2a,b, relevant to the observations (Table 1) are shown in Fig. 3.

At the high temperature end, the V = VO boundary is located at ~2080 K over a wide range of total pressures. Temperatures between the V = VO boundary and Ca-Ti-oxides and Ca-Al-oxide condensation curves should be characteristic for mid- to late M dwarfs because TiO and VO are major gases, and all neutral alkali atoms become more abundant. For these objects, the Li abundance derived from the Li I line is representative of the bulk Li abundance and can be used to decide an object's substellar nature.

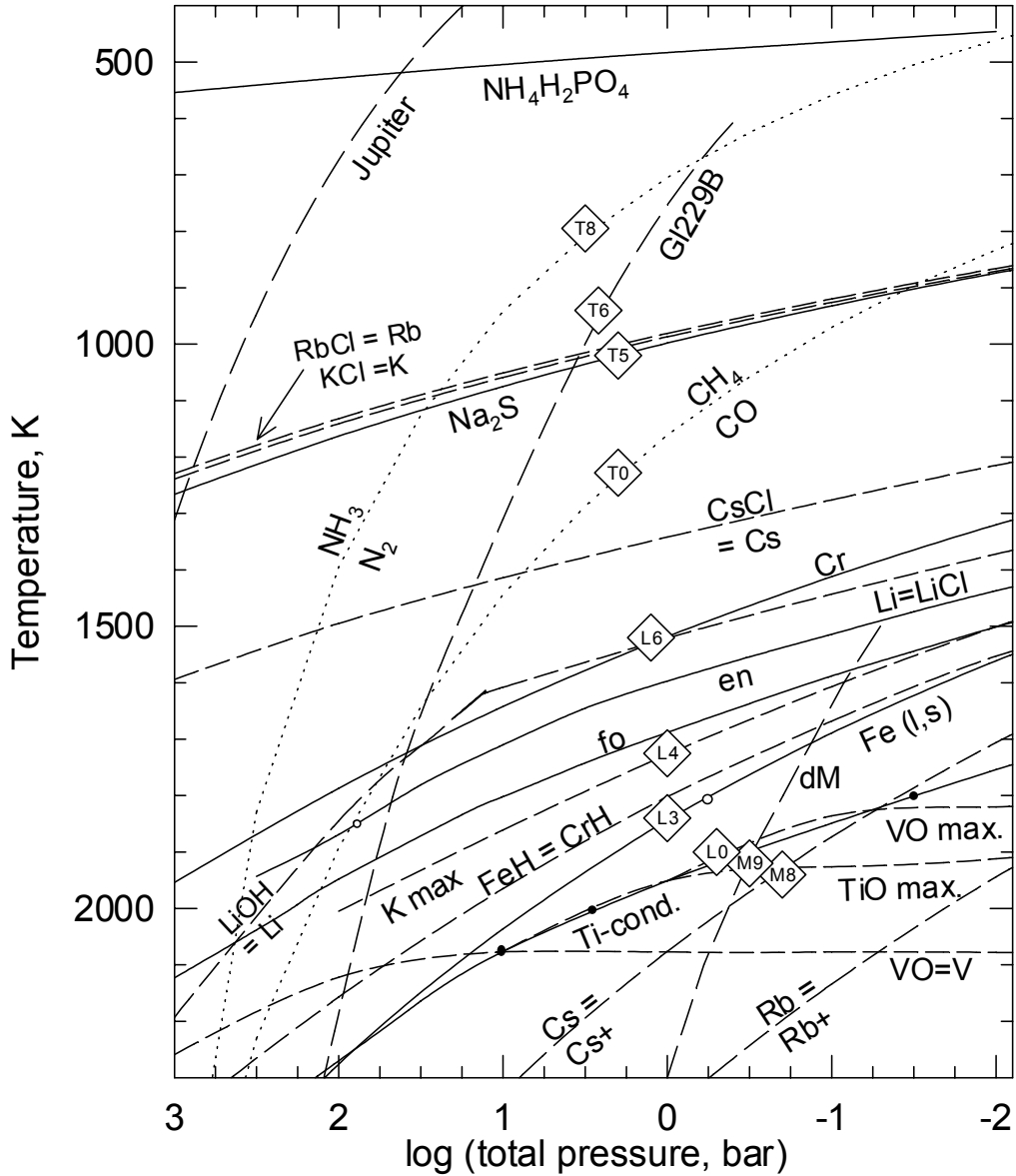

**Figure 3**. A summary of the major chemistry features from Figure 2a,b. The large diamonds indicate the temperature dependent chemical characteristics for a given spectral type. These "chemical" temperatures are only lower limits to the true effective temperatures. See the text for more explanations.

Chemical temperatures at the L/M dwarf transition using the TiO (g) and VO (g) abundances from thermochemical computations can be derived as follows (see Lodders 2002). In M8 dwarfs TiO bands are at maximum strength (K99, M99), hence the chemical temperature for M8 dwarfs are placed on the maximum TiO gas abundance curve in Fig.3. Similarly, VO bands peak in strength at M9, which places M9 onto the maximum VO gas curve. Total pressures should be lower than ~ 1bar because otherwise the maxima of TiO and VO both coincide with the Ca-Ti condensation curve. The position on the total pressure axis is further constrained by the positions of the $Cs = Cs^+$ and $Rb = Rb^+$ boundaries. The lines of monatomic Cs and Rb strengthen beyond M9 in part because the monatomic gases become the most abundant Cs and Rb bearing gases (Figs. 2b, 3). The chemical temperature marker for M8 is placed in the vicinity where the TiO max curve is crossed by the $Cs = Cs^+$ boundary (Fig. 3). This leads to temperatures of 1940 K for M8 and 1920 K for M9.

In L0, TiO and VO bands start to decline and temperatures must be below the Ti- and V-bearing condensate curve (labeled Ca-Ti cond.). Temperatures at the beginning of the L sequence are <1920K and are tentatively placed at 1900 K for L0. In addition to temperature, the TiO and VO abundances can also uniquely constrain the total atmospheric pressures at the M/L dwarf transition (see KL02).

A comparison with the relatively large range of estimates for the effective temperatures of 2000-2500 K at the M/L transition (Table 2) shows that the chemical temperature of ~ 1900 K is lower by 100-600 K.

The next major chemical change in the early L sequence is the condensation of metal, forsterite, and enstatite between 1840 and 1600 K (at 1 bar). Useful markers for the chemical temperatures are the curves for FeH=CrH and maximum K I abundances shown in Fig. 3. The K I lines peak in strengths and equal band strengths of FeH and CrH are seen in L4 dwarfs. At 1 bar, FeH=CrH occurs at 1800 K and the maximum in K abundance at 1725 K. For illustrative purposes, we place spectral type L3 onto the Fe condensation line since the Fe I lines disappear around L3 (Table 1) and type L4 onto the K maximum curve (Fig. 3). These temperatures are then also consistent with the position of the FeH=CrH abundance curve. As to be expected the chemical temperature range of 1725-1840 K for L3 and L4 is about 100 K lower than the effective temperatures of 1800-1950 K for L3 and L4 found by Vrba et al. (2004).

The next chemistry markers are at spectral type L6 where CrH bands start to weaken and which seems to be the lowest spectral type at which monatomic Li is still seen (Table 1). Interestingly, the condensation curve of Cr metal and the equal abundance curve for Li=LiCl are positioned very close together (Fig. 2, 3) and intersect at 1520 K (at ~1.3 bar), which is used as chemical temperature for L6. This temperature is only 50 K less than the effective temperature of 1570 K for L6 estimated by Vrba et al. (2004).

The chemical temperature at the L-T transition is related to the $CH_4$=CO boundary (Figs. 2, 3). In the total pressure range of 1.3 – 3 bar, the temperatures from the $CH_4$=CO boundary give 1180 – 1280 K. The lower bound in total pressure is that obtained for spectral type L6 above, and the upper bound comes from the positioning of the effective temperature of the T dwarf Gl229B in Fig. 3. We plot the chemical temperature for T0 at 1230 K (at 2 bar). This chemical temperature coincides with the effective temperature at the L-T transition given by Burgasser et al. (2002a), but is generally 100-200 K lower than other estimates of the effective temperatures around the L-T transition (Table 2). However, only knowledge of methane and CO *abundance ratios* in late L and T dwarfs are good temperature diagnostics (Lodders & Fegley 2002) but this requires quantitative CO and methane abundance determinations which in turn require high-resolution model spectra for analyses.

There are no other direct chemical markers for the higher spectral types within the T sequence. The re-occurrence and peak in strengths of the K I lines and FeH bands around T5 (see Table 1) is most plausibly explained by effects from cloud physics and not by chemistry. Burgasser et al. (2002b) suggested that the cloud deck becomes more patchy within the T sequence, which allows to see deeper into the atmosphere so that K I and FeH bands are detectable again. These spectral features then begin to fade at T6 (Table 1) and we note that the effective temperature of T6 is lower than the $Na_2S$ condensation temperature (see Fig. 3). Thus, the appearance of a new, small cloud deck from $Na_2S$ around spectral types T5/T6 can block again the view into deeper atmospheric layers and is probably partly responsible that FeH and K are no longer seen in the spectra of the late T dwarfs. If so, the $Na_2S$ condensation temperature of 1005-1035 (for $P_{tot}$ of 1.3-3 bar) may reflect the chemical temperature for T5. Finally, effective temperatures of 700-800 K at the end of the T sequence (Table 2) are close to the $NH_3 = N_2$ boundary at the 3 bar level (Fig. 3).

*Acknowledgements*. Work supported in part by NSF grant AST-0406963 and NASA Planetary Atmospheres Program grant NAG5-11958.